\documentclass[pdflatex,sn-mathphys-num]{sn-jnl}

\usepackage{graphicx}%
\usepackage{multirow}%
\usepackage{amsmath,amssymb,amsfonts}%
\usepackage{amsthm}%
\usepackage{mathrsfs}%
\usepackage[title]{appendix}%
\usepackage{xcolor}%
\usepackage{textcomp}%
\usepackage{manyfoot}%
\usepackage{booktabs}%
\usepackage{algorithm}%
\usepackage{algorithmicx}%
\usepackage{algpseudocode}%
\usepackage{listings}%
\usepackage{color}

\theoremstyle{thmstyleone}%
\theoremstyle{thmstyletwo}%

\theoremstyle{thmstylethree}%

\usepackage{xcolor}

\usepackage{cancel}

\begin{document}

\title[Article Title]{Hyperuniform patterns nucleated at low temperatures: Insight from vortex matter imaged in unprecedentedly large fields-of-view}


\author[1,2]{\fnm{Alexey} \sur{Cruz-Garc\'{i}a}}

\author[1,2]{\fnm{Joaquín} \sur{Puig}}

\author[3]{\fnm{Sergii} \sur{Pylypenko}}

\author[1,2]{\fnm{Gladys} \sur{Nieva}}

\author[4]{\fnm{Alain} \sur{Pautrat}}

\author[5]{\fnm{Alejandro Benedykt} \sur{Kolton}}

\author*[1,2]{\fnm{Yanina} \sur{Fasano}}\email{yanina.fasano@gmail.com}

\affil*[1]{\orgdiv{Low Temperatures Lab}, \orgname{Centro At\'{o}mico Bariloche}, \orgaddress{\street{CNEA},  \country{Argentina}}}

\affil[2]{\orgdiv{Instituto de Nanociencia y Nanotecnología}, \orgname{CONICET-CNEA}, \orgaddress{\street{Nodo Bariloche},  \country{Argentina}}}

\affil[3]{\orgdiv{Leibniz Institut for Solid State and Materials Research},  \orgaddress{\city{Dresden}, \country{Germany}}}

\affil[4]{\orgdiv{Laboratoire CRISMAT-EnsiCaen},  \orgaddress{\city{Caen},  \country{France}}}

\affil[5]{\orgdiv{Condensed Matter Theory Group}, \orgname{Centro At\'{o}mico Bariloche}, \orgaddress{\street{CNEA},  \country{Argentina}}}



\abstract{Hyperuniform patterns present enhanced physical properties that make them the new generation of 
cutting-edge technological devices. Synthesizing devices with tens of thousands of
components arranged in a hyperuniform fashion has thus become a breakthrough to achieve in order to implement these technologies. Here we provide evidence that extended two-dimensional hyperuniform patterns
spanning tens of thousands of components can be nucleated using as a template the low-temperature vortex structure obtained in pristine Bi$_{2}$Sr$_{2}$CaCu$_{2}$O$_{8}$ samples after following a field-cooling protocol.}

\keywords{vortex matter, superconductors, hyperuniformity}



\maketitle

\section{Introduction}\label{sec1}

Vortex matter in superconductors is a playground for studying how different types of disorder present in the host media, the superconducting sample,  shapes  
the nucleation of condensed matter phases with a broad spectrum of spatial correlations.~\cite{Blatter1994} 
In general, vortex phases nucleated in real samples present density fluctuations yielding a non-negligible variance of the  number of vortices $N$ enclosed within an in-plane area of radius $R$, $\sigma_{N}^{2} = \langle N^{2} \rangle - \langle N \rangle^{2}$. At one extreme of the statistical correlations lie ordered phases with quasi-long-range order.~\cite{Giamarchi1995,Klein2001,Fasano2005} In the other extreme lie disordered vortex systems which exhibit unbounded density fluctuations that grow faster than those 
of a point pattern generated by a uniform random distribution.~\cite{Roy2019,Llorens2020b,Puig2024}  Between these two extremes,  vortex phases exhibit an aperiodic in-plane arrangement of vortices  with density fluctuations increasing with distance more slowly than the studied area, namely with $\sigma_{N}^{2} \sim R^{\beta}$ with $0<\beta <2$. This results in vortices being more evenly spaced than those in a uniform random
distribution, asymptotically suppressing relative number fluctuations, $\sigma^2_N/\langle N \rangle \to 0$ in the limit of large window areas. These ubiquitous disordered vortex phases present a hidden order characterized by a slow down of density fluctuations at large wavelengths and belong to the structural class of hyperuniform systems.\cite{Gabrielli2003,Torquato2003,Torquato2018} Even though hyperuniformity is a long wavelength asymptotic property, experimental observations in different superconducting materials in moderate fields of view suggest some vortex phases are hyperuniform.~\cite{Rumi2019,Llorens2020,Puig2022,Aragon2023,Puig2023,Besana2024} Therefore, vortex matter can be used as a template to generate hyperuniform structures at low temperatures if nucleated in host media with particular disorder potentials.

Disordered hyperuniform patterns are universal structures present in different natural systems~\cite{Jiao2014,Dreyfus2015,Chen2018,Torquato2018,Rumi2019,Zheng2020,Llorens2020,Zheng2020,Nizam2021,Chen2021,Chieco2021,Zhang2022,Aragon2023,Philcox2023,MilorSalvalaglio2025} that  posses novel physical functionalities that make them exceptional for technological applications in comparison with conventional ordered materials.~\cite{Man2013,Zheng2020,Chen2023,Liang2024,Wang2025} For instance, a disordered hyperuniform network of Al$_{2}$O$_{3}$ walls and cylinders presents isotropic phononic and photonic bandgaps,~\cite{Man2013} thus blocking sound and light in all directions unlike crystals. Hyperuniform structures also posses enhanced thermal and electric transport properties, as well as mechanical resilience, outperforming conventional non-hyperuniform disordered media.~\cite{Chen2023}  Therefore, disordered hyperuniform systems are currently regarded as potential candidates for a new generation of technologies at the forefront of innovation. Most of these systems are disordered or class-II hyperuniform systems, a class of disordered structures with a moderate amount of density fluctuations but that retain the hidden hyperuniform order since $1<\beta<2$ for large wavelengths.~\cite{Torquato2018}

An algebraic increase of $\sigma_{N}^{2}$ is manifested in reciprocal space as an algebraic growth of the structure factor for short wave vectors $q$, namely $S(q) \propto q^{\alpha}$ when $q \rightarrow 0$. The relation between the growing exponents of $\sigma_{N}^{2}$ and $S(q)$ in a $d$-dimensional system is $\alpha= d - \beta$. Disordered or class-III hyperuniform systems present a structure factor growing with an exponent $0<\alpha < 1$. In contrast, ordered or class-I hyperuniform systems present $1<\alpha < 2$. Depending on the nature of the disorder of the host sample and the electromagnetic coupling of vortices with the material, ordered class-I~\cite{Rumi2019,Besana2024}  or disordered class-III~\cite{Llorens2020,Aragon2023} hyperuniform vortex phases can be nucleated at low temperatures. Then, vortex matter can be used not only as a template to generate hyperuniform patterns at low temperatures, but controlling its coupling with the disordered host media enables to obtain hyperuniform structures with tailored properties.

The reported hyperuniformity of vortex phases apparently challenges the fluctuation-compressibility theorem stating that systems with generic constituents at equilibrium with a thermal bath present a $S(q)$ in the $q \rightarrow 0$ limit proportional to the compressibility of the system.~\cite{PathriaBeale2011} The thermodynamic limit and the equivalence of ensembles, basic statistical physics concepts, stem from this simple scaling law. Therefore, theoretically, in equilibrium conditions only thermodynamically incompressible systems could present the hyperuniform hidden order at equilibrium,~\cite{Torquato2018} and strikingly vortex matter is a compressible elastic structure in three dimensions. 
In general, incompressibility at equilibrium is only accomplished when the interaction between constituents is repulsive and long-ranged, in contrast to vortex phases presenting typically short range interactions.~\cite{Torquato2018}
However, short-ranged interacting constituents can present hyperuniform arrangements for planes within a higher-dimensional system. This is indeed the case of the structure formed 
by the tips of superconducting vortices impinging
on the surface of the three-dimensional vortex structure nucleated in bulk samples
with point disorder.~\cite{Rumi2019,Llorens2020,Aragon2023,Besana2024}  Therefore, the reported hyperuniformity of vortex phases might not challenge the fluctuation-compressibility theorem after all.

In previous works we have shown that the hyperuniform correlations in the point pattern of vortices impinging in a plane arise from an effective long-range
interaction mediated by the elastic properties of
vortices along their length, namely across the sample thickness.~\cite{Rumi2019,Besana2024} Moreover, by means of  Langevin dynamics simulations of the quenching of the vortex structure on cooling we have shown that the hyperuniform order  progressively degrades on decreasing the magnitude of this effective long-range interaction as in the case of dramatically reducing the sample thickness.~\cite{CruzGarcia2025} These findings warn on the potentially negative impact of finite-size effects on
large-scale structural properties, which is crucial for designing
hyperuniform materials. Nevertheless,  the observation of hyperuniform vortex patterns at sufficiently thick samples is consistent
with the fluctuation-compressibility theorem since the density fluctuations of the vortex tips are associated with the compressibility of a single plane that
has a relatively large bulk tilting energy cost.

All the mentioned experimental evidence was  obtained from snapshots of structures frozen during cooling.~\cite{Fasano1999,CejasBolecek2016} Then the interpretation of these results raise the relevant question on whether the thickness dependence of hyperuniformity is an equilibrium effect, as predicted in Refs.~\cite{Rumi2019,Besana2024}, or rather an out-of-equilibrium effect arising from the slow dynamics during cooling.  Our recent simulation results indicate that finite size effects, particularly finite thickness effects, appear both at and out-of equilibrium conditions.~\cite{CruzGarcia2025}

For real samples with a bounded thickness $t$, at equilibrium conditions the discussed finite-thickness effect yields an in-plane crossover distance $l_{\rm fs}= (t/2\pi) \sqrt{c_{11}/c_{44}}$ above which the system is no longer hyperuniform.~\cite{Puig2022} 
Since hyperuniformity is a structural property in an asymptotic limit, ascertaining whether a bounded real system presents this hidden order or has reached this crossover in-plane distance requires direct imaging of the constituents of a system in extended fields-of-view. Here we study this issue by imaging vortices in a thick ($t \geq 20$\,$\mu$m) pristine Bi$_{2}$Sr$_{2}$CaCu$_{2}$O$_{8}$ in a very large field-of-view containing 33 000 vortices. Previous studies in this host media with weak random point disorder and for the same vortex density show that for a field-of-view of roughly 5000 vortices the vortex structure is ordered class-I hyperuniform with an exponent $\alpha = 1.3-1.5$ for different thick samples.~\cite{Rumi2019,Besana2024} Here we reveal that in the unprecedentedly large fields-of-view of up to 33 000 vortices the vortex structure nucleated in pristine Bi$_{2}$Sr$_{2}$CaCu$_{2}$O$_{8}$ samples with weak point disorder remains hyperuniform and the finite size crossover length $l_{\rm fs} > 180 a$.

\begin{figure}
    \centering
    \includegraphics[width=0.95\linewidth]{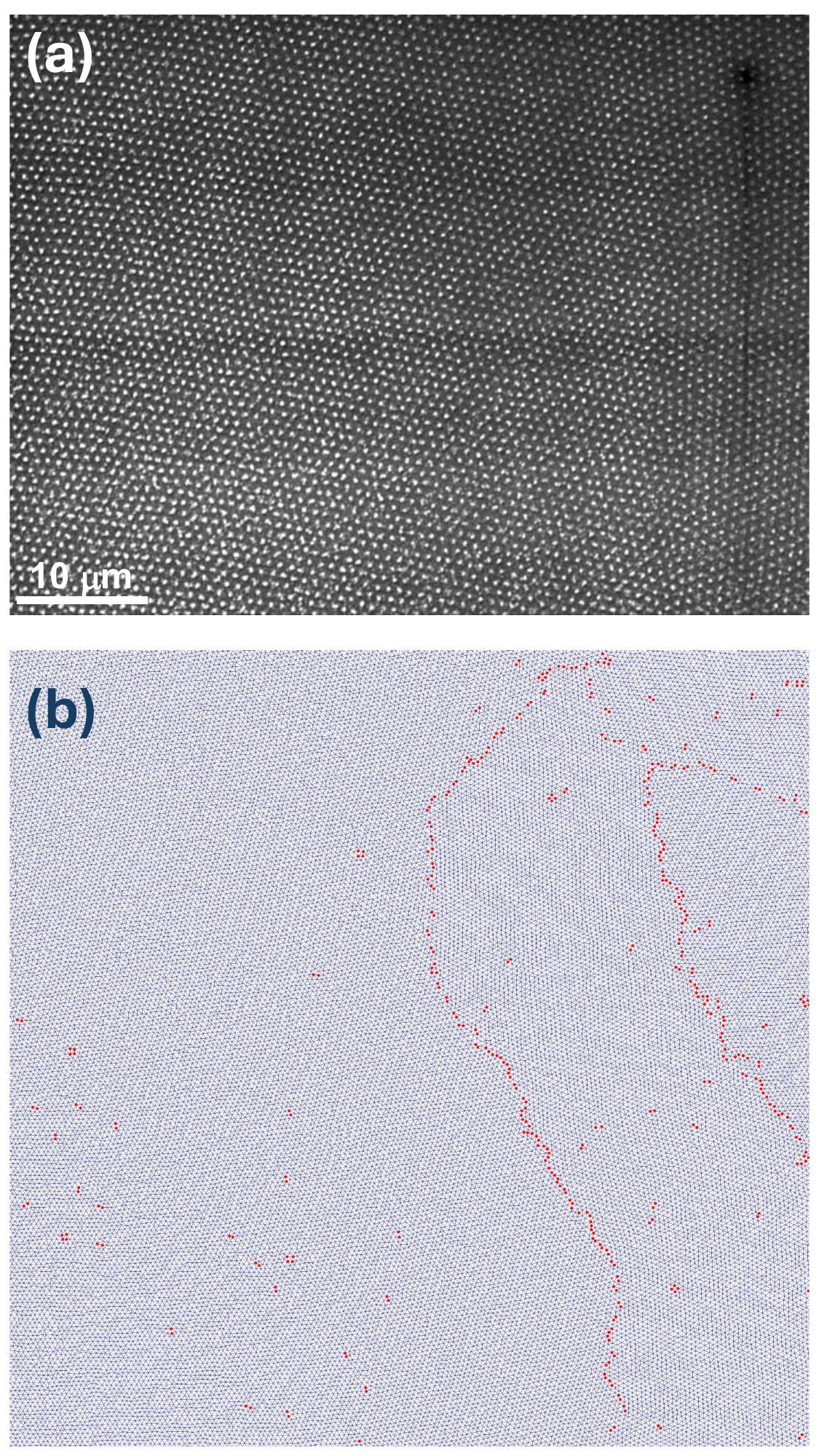}
    \caption{Structural properties of vortex matter nucleated at 30\,Oe in pristine Bi$_2$Sr$_2$CaCu$_2$O$_{8+\delta}$.(a) Zoom-in of a magnetic decoration of vortices in the sample with the largest surveyed field-of-view spanning 33 000 vortices. The zoom-in shows around 3 000 vortices imaged as white dots corresponding to the Fe clusters decorating vortex positions at the sample surface. (b) Delaunay triangulations of the largest studied field-of-view with 33 000 vortices in the same sample. Blue lines bond first neighbors and vortices highlighted in red are non-sixfold coordinated topological defects in the structure. }
    \label{Fig:figure1}
\end{figure}

\section{Experimental}\label{sec2}

We studied  pristine nearly-optimally doped Bi$_{2}$Sr$_{2}$CaCu$_{2}$O$_{8}$ samples with $T_{\rm c} \sim 90$\,K from two sources.  One batch of samples was grown by means of the flux
method~\cite{Correa2001} and on them we obtained vortex images spanning up to few thousands of vortices. 
Another batch grown also by the flux method containing much larger samples with tens of millimeter typical sizes~\cite{AragonSanchez2019} allowed us to obtain vortex images in large fields-of-view up to tens of thousands of vortices. All the studied samples are in the thick regime,~\cite{Besana2024}  presenting thicknesses larger than 20\,$\mu$m. The samples were specially selected  ensuring that they did not exhibit any visible planar defects as determined by magnetic decoration imaging.~\cite{Fasano2000,Puig2022} Then it is reasonable to assume that in the studied samples the dominant disorder is point-like, namely atomic-scale defects that arise randomly when the crystals are grown.

We apply the magnetic decoration  technique in order to image the positions of individual vortices on the surface of the sample covering extended fields-of-view, see for example the zoomed-in image of Fig.\,\ref{Fig:figure1} (a).
In a magnetic decoration experiment ferromagnetic particles evaporated at low temperatures are attracted towards the magnetic halo of vortices nucleated in superconducting samples in the mixed phase.~\cite{Fasano2008} At the magnetic core of vortices the local magnetic field presents a maximum that decays within a typical distance of the order of the  superconducting penetration depth, $\lambda(T)$.~\cite{Blatter1994}  In this way the ferromagnetic particles evaporated on the sample decorate the positions of the vortex tips impinging from the sample. Scanning electron microscopy is then used to obtain panoramic views of the vortex structure from the images of the sample surface containing ferromagnetic particles that decorate the vortex positions. Digitalizing the position of the ferromagnetic clusters allows the identification of vortex positions in extended fields of view.

 Magnetic decoration imaging is better suited to study vortex density fluctuations at large length scales than other imaging techniques which typically image only hundreds of vortices, such as scanning tunnelling spectroscopy~\cite{Petrovic2009,Suderow2014,Llorens2020}, magnetic force microscopy and scanning SQUID microscopy.~\cite{Llorens2020b} In addition, this technique can also be used to 
study the structural properties of vortex matter in extended fields-of-view nucleated on the same sample in different experimental realizations~\cite{Fasano1999,Fasano2000} and for different lengths of the vortex system.~\cite{Fasano2003}
The experimental protocol used in this work is a field cooling: We obtain snapshots of the vortex structures at low temperatures after cooling the system from the normal state under an applied field. The data presented in this work corresponds to vortices nucleated at a field of 30\,Oe and decorations performed at 4.2\,K.  While field cooling the vortex structure gets frozen, at lengthscales of the order of the lattice spacing $a$, at a temperature $\sim T_{\rm freez}$
intermediate between the first-order melting transition temperature and the decoration temperature.~\cite{Fasano1999} $T_{\rm freez}$ is
a characteristic temperature where the dynamics of the vortex structure is dramatically slowed down by the disorder of the host medium given by the pinning potential.~\cite{CruzGarcia2025} Thus this temperature is of the order of the irreversibility temperature at which pinning sets in, namely $T_{\rm freez} \sim T_{\rm irr} (B) \sim 0.9 T_{\rm c}$.~\cite{CejasBolecek2016,Dolz2014}

\section{Results}\label{sec3}

Figure\,\ref{Fig:figure1} (a) shows a zoom-in containing about 3 500 vortices from the largest field-of-view studied that includes approximately 33 000 vortices. Every vortex is imaged as a white dot that corresponds to the Fe clusters that decorate the position of vortices on impinging the sample surface. For the studied vortex density nucleated at $B=30$\,Oe the lattice spacing $a \sim 0.8$\,$\mu$m.
The vortex structure presents the quasi-long range positional order compatible with the Bragg glass phase~\cite{Kim1999,Fasano2005,Aragon2019} and some grain boundaries between very large grains. This is better appreciated in the Delaunay triangulation of panel (b) showing all vortices in the largest studied field-of-view. The Delaunay triangulation follows an algorithm to identify first neighbors allowing to study the coordination of each vortex. In the image  first neighbors are bonded with blue lines and non-sixfold coordinated vortices are highlighted in red. These topological defects form screw dislocations (a five-fold coordinated vortex adjacent to a seven-fold coordinated one) that appear isolated or grouped together in the boundaries separating large vortex grains with different orientations.

\begin{figure}
    \centering
    \includegraphics[width=0.8\linewidth]{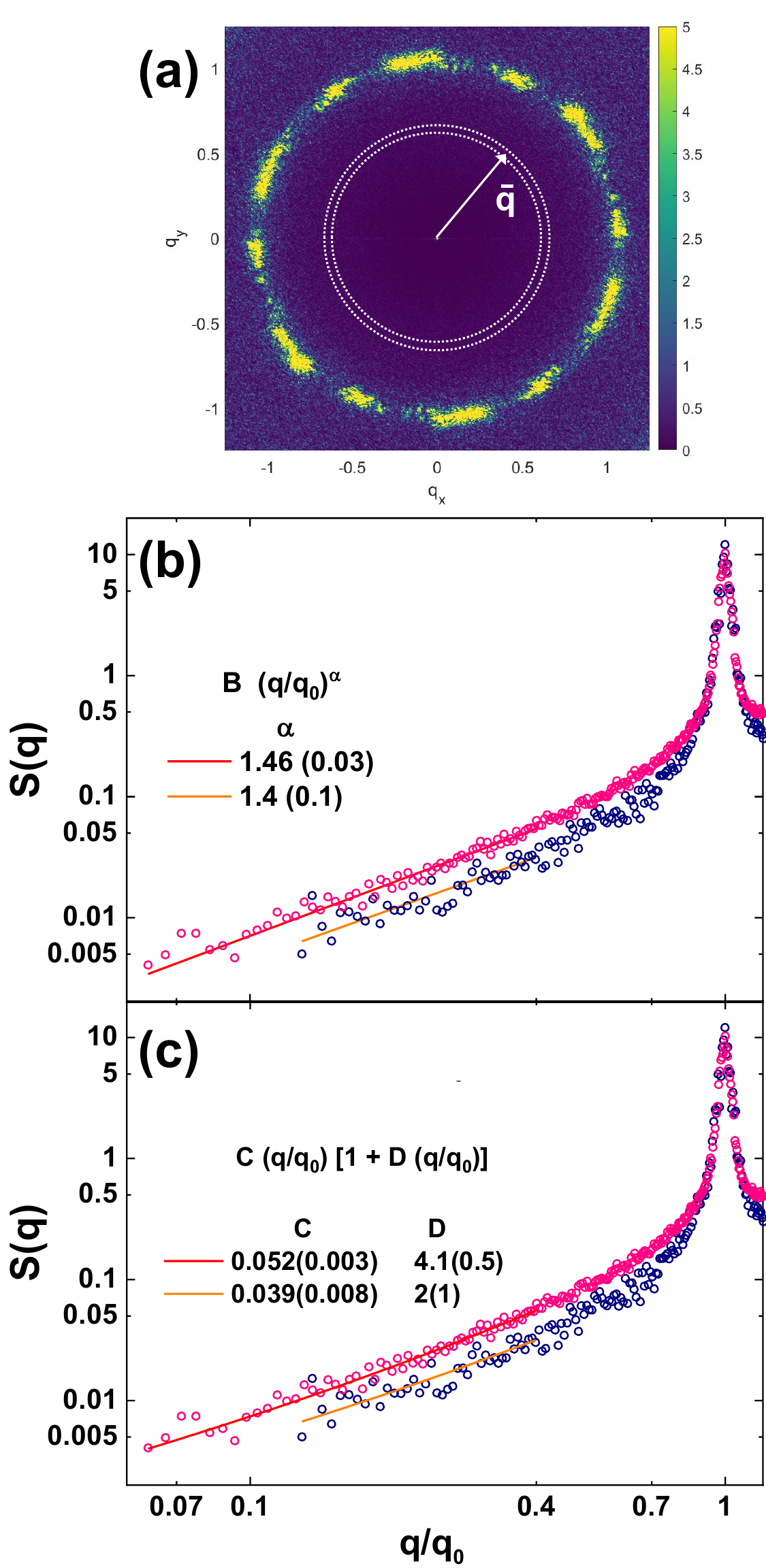}
    \caption{Structure factor data of vortex matter nucleated at 30\,Oe in pristine Bi$_2$Sr$_2$CaCu$_2$O$_{8+\delta}$.(a) Two-dimensional structure factor of the largest studied field-of-view computed after digitalizing the positions of roughly 33 000 vortices. Bragg spots (yellow features) appear at the Bragg wavevector $q_{0}$. The magnitude of $S(q_{\rm x},q_{\rm y})$ is  averaged along an infinitesimal circle of radius $q$, see dotted white lines, in order to obtain the angularly-averaged structure factor $S(q)$. (b)   $S(q)$ data for the smallest (blue open dots) and largest (pink open dots) studied fields-of-view 
    including 4 000 and 33 000 vortices, respectively. Red and orange lines are algebraic fits up to $q/q_{0} = 0.4$ yielding the exponents $\alpha$ indicated in the legend. (c) Same data fitted in the same $q/q_{0}$ range with the function $S(q) =C (q/q_0)\!\left(1 + D\,(q/q_0)\right)$ theoretically predicted for
    $\alpha=1$ type II   hyperuniformity with dispersive elastic constants (see text). Obtained fitting parameters with their errors indicated in the legend.}
    \label{Fig:figure2}
\end{figure}

Vortex positions are digitalized from the decoration images in order to obtain the structure factor by Fourier transforming the local density fluctuations of vortex positions, namely $S(q_{\rm x},q_{\rm y})=|\hat{\rho}(q_x,q_y,z=0)|^2$ at the sample surface, with $\rho(q_x,q_y,z)$ the vortex density matrix. Data of the two-dimensional structure factor obtained from the largest studied field-of-view are presented in Fig.\,\ref{Fig:figure2} (a). The nucleation of large grains in the structure is evident from the detection of several sextets of Bragg peaks at the Bragg wavevector $q_{0}$. From these data we compute the angularly-averaged structure factor $S(q)= S(\sqrt{q_x^2 + q_y^2})$ obtained by averaging the two-dimensional $S(q_x,q_y)$ data of an infinitesimal circle of radius $q$ over the polar angle, see schematics in Fig.\,\ref{Fig:figure2} (a). Angularly-averaged structure factor data for the smallest (4 000 vortices) and largest (30 000 vortices) studied fields-of-view are presented in  Fig.\,\ref{Fig:figure2} (b).

Irrespective of the size of the field-of-view, the log-log plot of Fig.\,\ref{Fig:figure2} (b) show that the structure factor decays algebraically in the $q \rightarrow 0$ limit. Fits to the data using $S(q) = B (q/q_{0})^\alpha$  yield $\alpha=1.46$ for the largest and $\alpha=1.4$ for the smallest fields-of-view. The prefactor $B$ is larger and the height of the Bragg peak is shorter for the structure nucleated in the sample where the largest field-of-view is studied, suggesting the magnitude of vortex density fluctuations is larger than for the structure nucleated in the another sample. This difference quite likely has origin in a difference in the magnitude of point disorder in both samples.
Regardless of these differences, the $\alpha$ exponents obtained in both cases are similar within their error and indicate that the studied vortex structure presents effective type-I hyperuniform properties as suggested in previous studies in smaller fields-of-view for various thick samples and vortex densities around 30\,G.~\cite{Rumi2019,Besana2024}

Strict type-II hyperuniformity with $\alpha = 1$---the behavior expected theoretically at equilibrium for interacting elastic lines in the absence of disorder, or for the vortex liquid at equilibrium \cite{Rumi2019}---cannot be ruled out. In the case of interacting vortices nucleated in a sample with negligible disorder, including dispersive corrections to the elastic moduli $c_{11}$ and $c_{44}$, leads to the modified small-$q$ limit expression for the structure factor,
$
S(q) \propto (q/q_0)\!\left(1 + D\,(q/q_0)\right),
$
with $D>0$.~\cite{Rumi2019} The fits of Fig.\,\ref{Fig:figure2} (c) shows that this expression also provides a reasonable description of the data with similar error than the algebraic fits of panel (b). This indicates that in the asymptotic $q\rightarrow 0$ limit type-II hyperuniformity considering dispersivity in the elastic constants is a viable alternative interpretation to characterize the structural properties of vortex matter nucleated in Bi$_2$Sr$_2$CaCu$_2$O$_{8+\delta}$ pristine samples.

Thus, for any of the two possible scenarios discussed above, vortex matter nucleated at 30\,Oe in pristine Bi$_2$Sr$_2$CaCu$_2$O$_{8+\delta}$ the vortex structure is hyperuniform up to lengthscales of $\sim 180 a$. Therefore, for this vortex system the finite size crossover length $l_{\rm fs} > 180 a$.

\section{Conclusion}

In conclusion, our work reveals that extended two-dimensional hyperuniform patterns spanning tens of thousands of components, can be nucleated at low temperatures when the host media where the structure is quenched on cooling exhibits uncorrelated weak point disorder. 
In the system we study here, vortex matter nucleated in pristine Bi$_{2}$Sr$_{2}$CaCu$_{2}$O$_{8}$ samples, the finite size crossover length  above which density fluctuations grow with an exponent close to system dimension  exceeds 180 lattice spacings. 
This vortex structure quenched in such a host media down to low temperatures can be used as a template to generate two-dimensional structural systems with strongly suppressed density fluctuations at large lengthscales. These results are important for designing a road-map to synthesize hyperuniform patterns for cutting-edge technological applications for large devices with tens of thousands of components.

\bmhead{Acknowledgements}

We acknowledge financial support from the organization of the 30th International Conference on Low Temperature Physics at Bilbao, Spain, in order for Y.F. to attend the conference. Y.F.
acknowledges funding from the Alexander von Humboldt
Foundation through the Georg Forster Research Award and
from the Technische Universität Dresden through the Dresden
Senior Fellowship Program. Work partially
supported by the National Council of Scientific and Technical
Research of Argentina (CONICET) through Grant No.
PIP 2021-1848 and by
the Universidad Nacional de Cuyo Research Grant No.
06/80020240100305UN.

\bibliography{biblio}
\end{document}